\documentclass{aa}  

\usepackage{graphicx}
\usepackage{xspace}
\usepackage{xcolor}     
\usepackage[breaklinks=true]{hyperref}
\usepackage{array}
\usepackage{soul}       
\usepackage{orcidlink}
\hypersetup{
    colorlinks,
    linkcolor={blue},
    citecolor={blue},
    urlcolor={blue}
}

\newcommand{\rcore}{r_{\rm core}}
\newcommand{\pa}{\phi}
\newcommand{\glee}{{\sc Glee}\xspace}

\newcommand{\p}{G165\xspace}

\usepackage{txfonts}
\graphicspath{{./}}
\begin{document}

\title{Strong lensing model and dust extinction maps\\of the host galaxy of type Ia supernova H0pe}

\titlerunning{Lens model of PLCK\,G165 and dust extinction from extended image modeling of SN H0pe host}

\author{
A.~Galan \inst{\ref{mpa},\ref{tum},\orcidlink{0000-0003-2547-9815}}\fnmsep\thanks{Corresponding author (\href{mailto:aymeric.galan@gmail.com}{aymeric.galan@gmail.com})}
\and
S.~Schuldt \inst{\ref{unimi},\ref{inafmilano},\orcidlink{0000-0003-2497-6334}}
\and
G.~B.~Caminha \inst{\ref{tum},\ref{mpa}}
\and
S.~H.~Suyu \inst{\ref{tum},\ref{mpa},\orcidlink{0000-0001-5568-6052}}
\and
R.~Ca\~{n}ameras \inst{\ref{lam},
\orcidlink{0000-0002-2468-5169}}
\and
S.~Ertl \inst{\ref{mpa},\ref{tum}}
\and
C.~Grillo \inst{\ref{unimi},\ref{inafmilano},\orcidlink{0000-0002-5926-7143}}
\and
A.~Acebron \inst{\ref{csic},\ref{inafmilano},\orcidlink{0000-0003-3108-9039}}
\and
B.~Frye \inst{\ref{uaz},\orcidlink{0000-0003-1625-8009}}
\and
A.~M.~Koekemoer \inst{\ref{stsci},\orcidlink{0000-0002-6610-2048}}
\and
M.~Pascale \inst{\ref{ucla},\orcidlink{0000-0002-2282-8795}}
\and
R.~Windhorst \inst{\ref{asu}},
\orcidlink{0000-0001-8156-6281}
\and
J.~M. Diego \inst{\ref{ifca},\orcidlink{0000-0001-9065-3926}}
\and
N.~Foo \inst{\ref{asu},\orcidlink{0000-0002-7460-8460}}
}
\institute{
Max-Planck-Institut f\"ur Astrophysik, Karl-Schwarzschild-Str. 1, D-85748 Garching, Germany \label{mpa}
\and 
Technical University of Munich, TUM School of Natural Sciences, Physics Department,  James-Franck-Stra{\ss}e 1, 85748 Garching, Germany
\label{tum}
\and 
Dipartimento di Fisica, Universit\`a degli Studi di Milano, via Celoria 16, I-20133 Milano, Italy
\label{unimi}
\and
INAF -- IASF Milano, via A.~Corti 12, I-20133 Milano, Italy
\label{inafmilano}
\and
Aix-Marseille Université, CNRS, CNES, LAM, Marseille, France
\label{lam} 
\and
Instituto de Física de Cantabria (CSIC-UC), Avda. Los Castros s/n, 39005 Santander, Spain
\label{csic}
\and
Department of Astronomy/Steward Observatory, University of Arizona, 933 N. Cherry Avenue, Tucson, AZ 85721, USA
\label{uaz}
\and
Space Telescope Science Institute, 3700 San Martin Drive, Baltimore, MD 21218, USA
\label{stsci}
\and
Department of Physics \& Astronomy, University of California, Los Angeles, 430 Portola Plaza, Los Angeles, CA 90095, USA
\label{ucla}
\and
School of Earth and Space Exploration, Arizona State University, Tempe, AZ 85287-6004, USA
\label{asu}
\and
Instituto de F\'isica de Cantabria (CSIC-UC). Avda. Los Castros s/n. 39005 Santander, Spain
\label{ifca}
}
 
\abstract{
Strong gravitational lensing by massive galaxy clusters offers rare opportunities to observe multiple images of distant ($z\gtrsim2$) Type Ia supernovae (SNe) and to resolve the properties of their host galaxies. A recent outstanding example is the Type Ia SN H0pe ($z=1.78$), which the James Webb Space Telescope (JWST) discovered in NIRCam images, when the galaxy cluster PLCK\,G165.7+67.0 (\p, $z=0.35$) still produced three images of it. In this work, we build a new strong lensing model of \p, first using only the positions of multiple images of background galaxies. We then significantly increase the number of constraints around the position of SN H0pe by modeling the extended surface brightness of the SN host galaxy. Including extended image information reduces the average uncertainty on mass model parameters by more than an order of magnitude. We also study the spatial distribution of dust in the arc to estimate the dust extinction at the position of SN H0pe. We find good statistical agreement of the extinction estimate, at $\lesssim1\sigma$, with three fully independent methods based on spectral energy distribution fitting. Moreover, our extended-image lens model of \p allows us to map the dust distribution of the host galaxy from the image plane to the source plane. Supernova H0pe exploded in a region with a relatively high extinction ($A_V \approx 0.9\ {\rm mag}$) at around $\sim 1\ {\rm kpc}$ from its host center. This work shows that extended image modeling in lensing clusters simultaneously reduces the uncertainty on lens model parameters and enables spatially resolved analyses of lensed transients' host galaxies. Such modeling advances are expected to play an important role in future cosmological analyses using strongly lensed SNe.
}


\maketitle

\section{Introduction \label{sec:introduction}}

The recently discovered Type Ia supernova (SN) H0pe \citep{2023TNSAN..96....1F,2024ApJ...961..171F}, which was triply imaged by the galaxy cluster PLCK\,G165.7$+$67.0 (hereafter \p) is exceptional in many respects. This system offers a rare opportunity to measure the Hubble constant ($H_0$) jointly using the standard candle nature of Type Ia SNe and the time delays between the multiple images \citep{2025ApJ...979...13P}. The host galaxy of SN H0pe appears as a highly magnified giant arc, providing unique insights into the properties of Type Ia SN hosts at redshift $\gtrsim 2$.

\begin{figure*}
\centering
  \includegraphics[width = 1.0\textwidth]{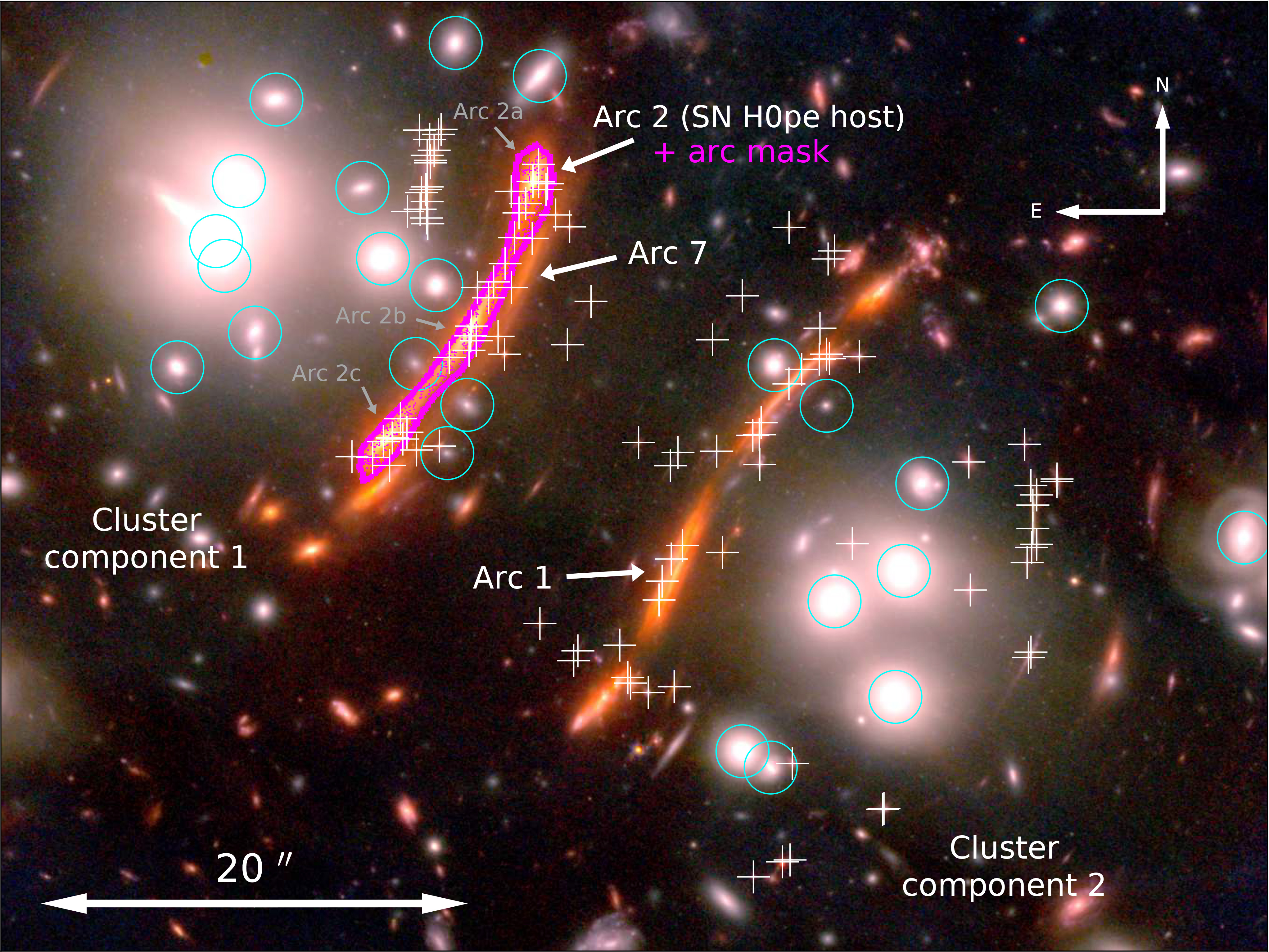}
  \caption{JWST color-composite image of \p (where the combination for blue is $\rm F090W + F115W + F150W$; green is $\rm F200W + F277W$; and red is $\rm F356W + F410M + F444W$). White crosses indicate the position of point-like multiple images used as model constraints \citep[presented in][]{2024ApJ...961..171F}. The magenta contour outlines the mask used to model the surface brightness of the SN H0pe host (Arc 2, composed of the SN host images 2a, 2b, and 2c). Cyan circles show a subset of the cluster members included in our lens models. Our lens models also include two cluster-scale mass components, each located around the main concentrations of cluster members. The figure also indicates Arc 1 and Arc 7, which we mention in the text.}
  \label{fig:cluster_multiple_images}
\end{figure*}

The massive galaxy cluster \p ($z=0.348$) was discovered as part of Planck's Dusty Gravitationally Enhanced subMillimeter Sources follow-up program, thanks to the extremely bright far-infrared emission from a background lensed, dusty star-forming galaxy \citep{2015A&A...582A..30P,2015A&A...581A.105C,2016MNRAS.458.4383H}. \p attracted attention for its numerous giant arcs, one of which (usually referred to as Arc 1; see Fig.~\ref{fig:cluster_multiple_images}) has been studied in detail, first mainly at submillimeter wavelengths, then at optical and near-infrared wavelengths \citep{2018A&A...620A..60C,2019ApJ...871...51F}. The location of critical lines from several strong lensing models of \p reveals high-magnification lensing events in different arcs \citep[e.g.,][]{2018A&A...620A..60C,2019ApJ...871...51F,2022ApJ...932...85P,2024ApJ...961..171F,2024ApJ...973...25K}, potentially indicating new ones in future observations. 
The James Webb Space Telescope (JWST) program Prime Extragalactic Areas for Reionization and Lensing Science \citep[PEARLS;][]{Windhorst2023} therefore included \p in its sample to shed further light on the dynamical (likely merging) state of the cluster and to increase the number of known strongly lensed sources. 

The first images from the JWST Near Infrared Camera (NIRCam) from PEARLS revealed the three images of SN H0pe ($z=1.78$), confirmed to be Type Ia with JWST Near Infrared Spectrograph (NIRSpec) follow-up, as detailed in \citet{2024ApJ...961..171F}. Its host galaxy (referred to as Arc 2; see Fig.~\ref{fig:cluster_multiple_images}) is a moderately dusty star-forming galaxy, possibly a member of a larger galaxy group \citep{2024ApJ...961..171F}. The time delays between the SN images were measured using two approaches. \citet{2024ApJ...967...50P} estimated the delay by reconstructing the SN light curves from multi-epoch JWST/NIRCam photometry, using the complete surface brightness reconstruction of Arc 2 from this work. \citet{2024ApJ...970..102C} used JWST/NIRSpec spectroscopy to determine the SN phase via template fitting, from which they inferred the time delays. In both approaches, the absolute magnifications were measured for each of the three SN appearances and used for the $H_0$ inference \citep{2025ApJ...979...13P}.

Regardless of the goal, whether measuring $H_0$ from the time delays or studying the SN host galaxy, an accurate lens model of the foreground deflector must first be constructed. Recently, \citet{2025ApJ...979...13P} brought together seven independent teams to perform the strong lens modeling of \p. The same set of cluster members, additional galaxies, and lensing constraints was used by all teams. Although all models showed generally good fits, the separation between the predicted and observed positions of the model in all images varied significantly between them (from $0\farcs07$ to $1\farcs1$ in rms). Such a spread, given the same set of observables, should not be surprising due to the variety of modeling assumptions: some models are based on varying numbers of parametric profiles, others on regularized mass-density grids, and some combine both approaches. Despite the variety in modeling approaches, \citet{Agrawal2025} found that all these seven models of \p systematically overestimate the magnification of SN H0pe, potentially biasing the inferred $H_0$ value toward higher values. With additional and improved lensing constraints (more multiple-image families, more accurate astrometry, and redshifts), and provided that models are flexible enough, one may expect that different approaches will produce more consistent results. This hypothesis was recently explored by \citet{2025OJAp....8E..37P}, although the type of lensing constraints remained limited to point-like images, as in most cluster-scale lens models.

Some analyses have included spatially extended constraints by explicitly modeling the surface-brightness distribution of the arcs. In a galaxy-scale, cluster-hosted lens system, \citet{2024A&A...689A.304G} showed that explicitly modeling the lensed source surface brightness provides meaningful corrections to the original cluster lens model. At the scale of galaxy groups, extended arcs have been used in several analyses, such as \citet{2010A&A...524A..94S}, \citet{2022A&A...668A.162W}, \citet{2023A&A...671A..60B}, \citet{2024ApJ...973....3S}, and \citet{2025RAA....25f5013D}, to significantly better constrain the mass distribution in lensing groups. The reconstruction of extended arcs on cluster scales is less common in the literature, particularly due to inherently larger computational costs. However, this long-standing problem is now being addressed. For instance, \citet{2024ApJ...976..110A} performed pixelated source modeling of a quasar host galaxy triply lensed by a galaxy cluster, providing new insights into quasar-host relationships at high redshift while significantly improving the lens model precision. In the galaxy cluster hosting the multiply imaged SN Refsdal \citep{2015Sci...347.1123K}, Schuldt et al. (in prep.) also modeled the full surface brightness of the SN host, finding one to two orders of magnitude improvements in mass-model parameter uncertainties.

Studying the hosts of Type Ia SNe at redshifts $z \gtrsim 2$ is key to understanding their joint evolution over time. Host properties such as stellar mass, star-formation rate, and metallicity can all influence the SN luminosities and light-curve shapes \citep[see][for a recent review]{2025A&ARv..33....1R}. In particular, extinction by dust in the vicinity of SNe and along the line of sight directly affects their spectral energy distribution, potentially introducing biases into distance measurements \citep[e.g.,][]{2010ApJ...715..743K,2024MNRAS.534.2263P,2023MNRAS.518.1985M}. Therefore, at $z \sim 2$, where galaxies are younger and more metal-poor, such studies are especially important to ensure the reliability of SNe Ia as standard candles. Additionally, constraining the relationship between SN brightness and the size and mass of their host is important to avoid biases in magnitude-based selections of SNe \citep[e.g.,][]{2018A&A...615A..68R,2020ApJ...901..143U}, a challenge further complicated by lensing magnification for lensed SNe. Studying the environment of the most distant SNe, often revealed by massive strong lenses such as clusters, is also key to understanding the age and composition of SN progenitors and to better understand the triggering processes of Type Ia explosions and how their rates evolve over time \citep[e.g.,][]{2020MNRAS.499.1424H,2024ApJ...969...80C}. Specifically for SN H0pe, \citet{2024ApJ...961..171F} presented a NIRCam+NIRSpec spectrophotometric analysis to extract star-formation properties and quantify the color excess in Arc 2, while \citet{2024ApJ...970..102C} measured dust extinction parameters directly from the NIRSpec spectra.

In the present work, we follow an approach similar to that of \citet{2024ApJ...976..110A}, using the \glee modeling code \citep{2010A&A...524A..94S, Suyu+2012} to incorporate the entire Arc 2 from the SN H0pe host into the set of lensing constraints. We discuss the precision improvements in the properties of the two cluster-scale, dark-matter-dominated components of this merging cluster. Our extended-source model allows us to reconstruct a multi-band image of the SN H0pe host galaxy. Finally, we provide dust extinction maps in both image and source planes of that galaxy, revealing the dust environment of SN H0pe, and estimate the dust extinction at the SN position. We also compare our results with those of previous analyses \citep{2024ApJ...961..171F,2024ApJ...967...50P,2024ApJ...970..102C}.

When required, we assume a flat $\Lambda$ cold dark matter cosmology, with $H_0 = 70\ {\rm km}\,{\rm s}^{-1}\,{\rm Mpc}^{-1}$, $\Omega_{\rm m} = 0.3$, and $\Omega_{\rm \Lambda} = 0.7$. This cosmology yields angular sizes of $4.9\ {\rm kpc\, arcsec}^{-1}$ at $z=0.348$ (cluster plane) and $8.4\ {\rm kpc\, arcsec}^{-1}$ at $z=1.78$ (SN host plane).

This paper is organized as follows. We describe the imaging data of \p in Sect.~\ref{sec:data}. We present the results of our point-like-only lens model in Sect.~\ref{ssec:SLmod:pointlike}, while we show the results of our improved model, including the reconstruction of the SN host, in Sect.~\ref{ssec:SLmod:esr}. We focus on the properties of the host galaxy and build its dust extinction map in Sect.~\ref{sec:sn_host}. Finally, we conclude our work in Sect.~\ref{sec:conclusions}.

\section{Data \label{sec:data}}

The data acquisition and reduction are fully described  in \citet{2024ApJ...961..171F}; we provide a short summary here. The JWST observed \p under the PEARLS program (ID 1176, P.I. R.~Windhorst) in March 2023 with NIRCam.
Additional NIRCam imaging and NIRSpec multi-object spectroscopy were obtained in a Director's Discretionary Time program (PID 4446, P.I. B.~Frye) in April and May 2023. The exposure times range from $\approx 0.7$~hours to $1.4$~hours in eight filters in the short- and long-wavelength channels. After calibration with the JWST pipeline\footnote{\url{https://github.com/spacetelescope/jwst}} \citep{2023zndo...6984365B}, we combined the images into mosaics following the approaches first described by \cite{2011ApJS..197...36K}, updated for JWST \citep[see][section 2 and Table 1 for additional details]{2024ApJ...961..171F}.
Figure~\ref{fig:cluster_multiple_images} shows the NIRCam color-composite image of the central region of \p. We used the original PEARLS imaging data (epoch 1), which have longer exposure times than subsequent epochs\footnote{Epochs 2 and 3 (PID 4446, P.I. B.~Frye).}. The recently obtained template images without the SN were not available at the time of this analysis\footnote{Epochs 4 and 5 (PID 4744, P.Is B.~Frye, J.~Pierel). These template images will be presented in Agrawal et al. (in prep.).}. We used the same reduced data as in \citet{2025ApJ...979...13P}.

Spectroscopic redshifts were measured with NIRSpec and additional ground-based observations  \citep[for more details, see][]{2024ApJ...961..171F}, bringing the total number of families with measured spectroscopic redshifts to five.
Of the cluster members, 35 are spectroscopically confirmed at $z=[0.329,\, 0.368]$, which corresponds to a peculiar velocity of $4\,000\ {\rm km\,s}^{-1}$ relative to the mean cluster redshift $z_{\rm d}=0.348$. We used this sample of cluster galaxies to determine the loci in color-color space and to select an additional 130 cluster members using a photometric approach, down to magnitude $m_{\rm F150W} = 23.5$. Overall, we used the same 165 cluster members as in the lens models of\citet{2025ApJ...979...13P}. This catalog includes the 161 cluster members of \citet{2024ApJ...961..171F} plus four additional galaxies that either lack spectroscopic redshifts, lie near the cluster's redshift, or are far from the cluster's centroid \footnote{M.~Pascale, private communication; see also \citet{2025ApJ...979...13P}.}.

\section{Strong lens modeling of the galaxy cluster \p}

\subsection{Lens model with point-like image positions \label{ssec:SLmod:pointlike}}

To describe the lensing effect of \p, we assume two main mass components: one that represents the cluster-scale mass distribution and accounts mainly for dark matter, and a second that describes the small-scale galaxies that belong to the cluster. This merging cluster, being clearly bimodal, is itself composed of two components. Since we focus on constraining our models at the location of the two main cluster components, we omit external shear and other large-scale components far outside the field of view of Fig.~\ref{fig:cluster_multiple_images}.

For the cluster-scale component, we adopt the pseudo-isothermal elliptical mass distributions \citep[PIEMD;][]{1993ApJ...417..450K}, whose dimensionless projected mass distribution for a source at $z_{\rm s}\rightarrow \infty$ is given by
\begin{equation}
\label{eq:piemd}
\kappa(R) = \frac{\theta_{\rm E}}{2\sqrt{R(x,y, \varepsilon)^2 + \rcore^2}},
\end{equation}
where $\theta_{\rm E}$ is the Einstein radius (i.e., the mass normalization parameter) and $\rcore$ is the core radius.
The quantity $R$ is constant over ellipses with ellipticity $\varepsilon$, and is given by $R^2 = {(x-x_{\rm c})^2}/{(1+\varepsilon)^2} + {(y-y_{\rm c})^2}/{{(1-\varepsilon)^2}}$, where
$x$ and $y$ are along the semi-major and semi-minor axes, and $(x_{\rm c}, y_{\rm c})$ is the centroid of the profile.  The profile is then rotated by a position angle $\pa$. Our position angles are measured counterclockwise from the positive $x$ axis. Therefore, the PIEMD mass profile is fully described by six parameters. The initial positions of the two PIEMD profiles were located around the two main clumps of galaxy members (see Fig.~\ref{fig:cluster_multiple_images}).

In addition to the smooth cluster mass component, we account for the mass distribution of individual galaxies, some of which are indicated as cyan circles in Fig.~\ref{fig:cluster_multiple_images}. These galaxies belong mainly to the 161 selected cluster members of \p at redshift $z=0.348$. We also include the four additional galaxies present in the catalog of \citet{2025ApJ...979...13P} at the same redshift, given the current uncertainties regarding their nature (we defer testing the effect of this assumption to future models). Each galaxy in our model is parameterized by a circular dual pseudo-isothermal mass density \citep[dPIE; see][]{2007arXiv0710.5636E, 2010A&A...524A..94S}, with convergence given by
\begin{equation}
\label{eq:dpie}
\kappa_i(R) = \frac{\theta_{{\rm E},i}}{2} \left( \frac{1}{R(x,y, \varepsilon)} - \frac{1}{\sqrt{R(x,y, \varepsilon)^2 - r^2_{{\rm cut},i}}} \right),
\end{equation}
where $r_{\rm cut,i}$ is the cut radius of cluster member $i$, and $R$ is the radial coordinate (with elliptical coordinates defined as in Eq.~\ref{eq:piemd}). As there are not enough constraints to optimize the parameters of all cluster members individually, we reduced the total number of free parameters by making use of the information from their light distributions. In particular, we fixed the centroid of each galaxy to its observed light. We further reduced the number of parameters by assuming scaling relations similar to the Faber–Jackson relation for elliptical galaxies \citep[e.g.,][]{2007NJPh....9..447J,2016ApJ...822...78G}:
\begin{equation}
\theta_{{\rm E},i} = \theta_{\rm E,ref}\left( \frac{L_i}{L_\star} \right)^{0.5},\; r_{{\rm cut},i} = r_{\rm cut,ref} \left( \frac{L_i}{L_\star} \right)^{0.5}.
\label{eq:scaling_relation}
\end{equation}
Specifically, the relationship $\theta_{{\rm E},i}\propto L_i^{0.5}$ assumes a constant mass-to-light ratio in cluster members. Our reference galaxy ( $\theta_{\rm E}=\theta_{\rm E, ref}$ and $r_{\rm cut}=r_{\rm cut, ref}$) is a bright cluster member of \p located at (RA, Dec) = (11:27:06.6969, +42:27:50.375), which is outside the field of view of Fig.~\ref{fig:cluster_multiple_images} (toward the west). 

We used  41 image families as lensing constraints, leading to 106 point-like images. A subset of 15 image families do not have spectroscopic redshifts; thus, we leave their redshift free to vary (with uniform priors between 1 and 10). We note that eight of the 106 multiple images are located in the SN H0pe arc.

Our model contains 165 cluster members linked through the two parameters of Eq.~\ref{eq:scaling_relation} ($\theta_{\rm E, ref}$ and $r_{\rm cut, ref}$),  in addition to the two cluster-scale components with six parameters each, resulting in a total of $N_{\rm free}^{\rm mass} = 2 + 2 \times 6 = 14$ free mass model parameters. The 41 image families yield $N_{\rm free}^{\rm src-pt} = 2 \times 41 + 15 = 97$ free source parameters (two coordinates per source). The 106 observed point-like positions provide $N_{\rm con}^{\rm img} = 2 \times 106 = 212$ constraints. Therefore, our position-based lens model has $N_{\rm dof}^{\rm pos} = N_{\rm con}^{\rm img} - ( N_{\rm free}^{\rm mass} + N_{\rm free}^{\rm src-pt} ) = 101$ degrees of freedom.

\begin{figure*}
  \includegraphics[width = 1\columnwidth]{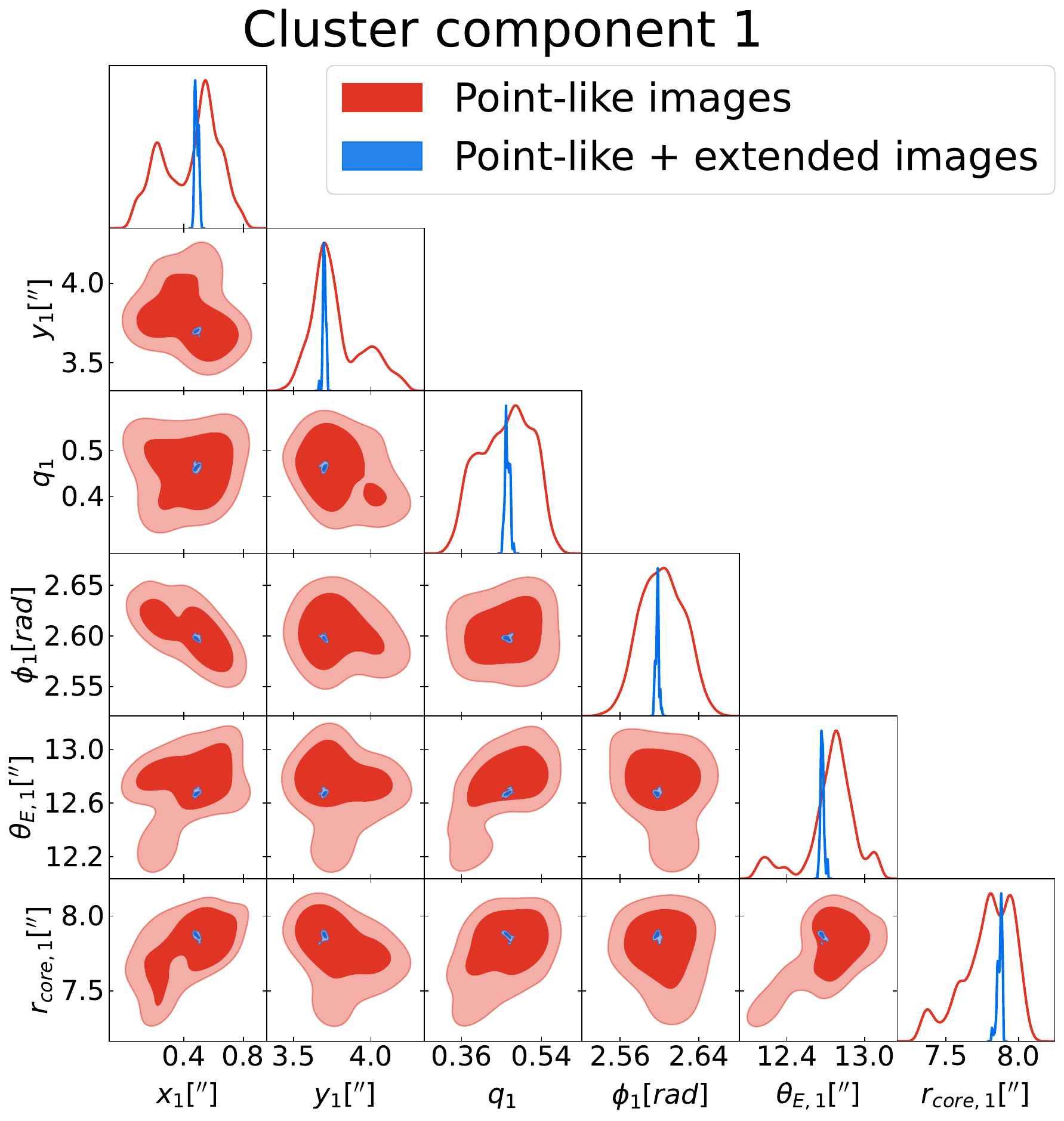}
  \includegraphics[width = 1\columnwidth]{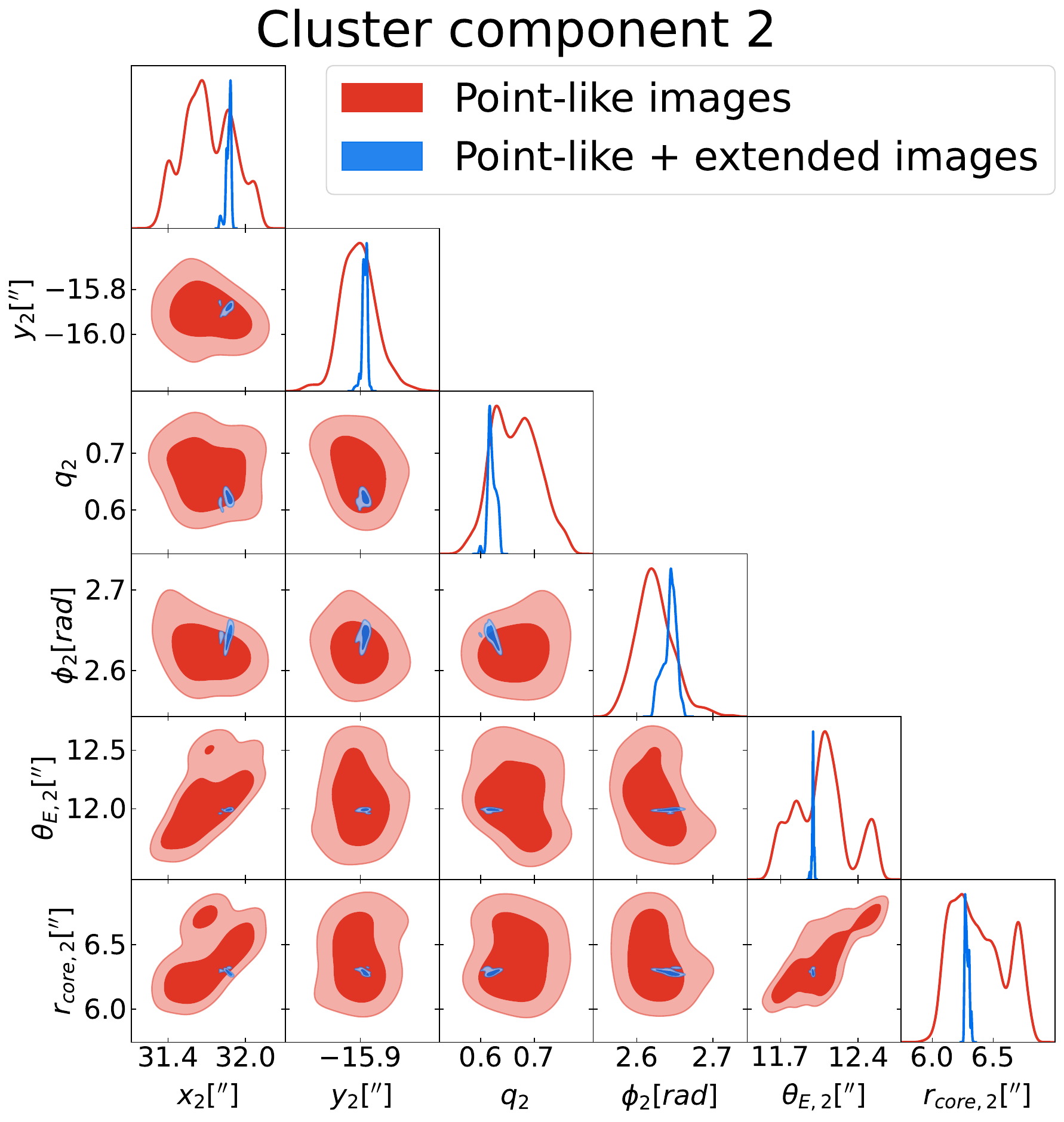}
  \caption{Joint posterior distribution of the mass parameters for the two cluster-scale components. Contours show the 68\% and 95\% confidence levels for the model using only the point-like images as constraints (red) and for the model including the full surface brightness of SN H0pe host galaxy (blue).}
  \label{fig:corner_mass}
\end{figure*}

\subsection{Lens model with point-like and spatially extended images of the SN host galaxy \label{ssec:SLmod:esr}}

Building on our previous model, we incorporated the full surface brightness distributions of the SN host galaxy (Arc 2) as an additional constraint alongside most of the point-like constraints from Sect.~\ref{ssec:SLmod:pointlike}. Specifically, we used the pixel intensity values of the lensed SN host galaxy as constraints. We thus excluded the eight point-like image families located within the arc mask (magenta contour in Fig.~\ref{fig:cluster_multiple_images}), yielding 84 instead of 106 point-like constraints in this extended-image lens model\footnote{The full catalog of cluster members and lensing constraints used in our lens models are available at \url{https://github.com/aymgal/SN_H0pe_extended_modeling}.}. With this approach, we reduced the model uncertainties due to the additional constraints from the light distribution of Arc 2 and, at the same time, reconstruct the morphology of the SN H0pe host galaxy.

We used the \glee \citep{2010A&A...524A..94S, Suyu+2012} software because of its capability to model extended images of gravitational arcs, not only in galaxy-scale systems \citep[e.g.,][]{Suyu+2013, Wong+2017, Ertl+2023}, but also on group and cluster scales \citep[e.g., ][]{2022A&A...668A.162W, 2023A&A...671A..60B,2024ApJ...976..110A}. The mass modeling method using spatially extended images of lensed galaxies is described in \citet{2010A&A...524A..94S}, and here we summarize its main aspects.

We adopted the same parametrizations for the lens mass distribution (smooth cluster halo and the individual cluster galaxies) as described in the previous section. For a given set of values for these parameters, we computed the deflection angles at each image pixel. Given the lensing deflection field and using the pixels in the arc mask, their associated errors, and a model of the point spread function (PSF), we obtain the optimal reconstruction of the SN host galaxy on a regular grid of pixels in the source plane of the SN through a linear matrix inversion described in detail in \citet{2006MNRAS.371..983S}. To construct the PSF model, we used three stars in the field and followed the same procedure described in \citet{Ertl+2023}. Our per-pixel error map is the quadratic sum of contributions from Poisson noise and background noise, obtained from the science data and the inverse variance weight maps, respectively. We used curvature regularization to invert the pixelated source, which reduces the curvature (second derivatives) in the source intensity grid to avoid overfitting the noise in the data. We then mapped this source intensity distribution to the image plane using the deflection field and the PSF model to obtain the reconstructed lensing arc, and we compared it directly to the observed intensity pixel values inside the arc mask. Based on the difference between the reconstructed and observed arcs, we compute the likelihood of the extended image of Arc 2.

We used the F200W filter to model Arc 2 because it provides the best compromise between the signal-to-noise ratio (S/N) of the extended arc and the PSF size of JWST. Because the SN extent is unresolved, our extended-source model should not attempt to reconstruct their flux on the source plane. We therefore increased the data uncertainties by a factor of $\sim 10^3$ within circular regions of four pixels in radius, centered on each SN image, effectively excluding these pixels from the data likelihood term. Compared to the position-based model of Sect.~\ref{ssec:SLmod:pointlike}, the extended-image model includes additional parameters and constraints. The number of additional free parameters equals the effective number of source pixels, which in our case is $N_{\rm free}^{\rm src-ext}\approx135$ \citep[for details of its computation, see][]{2006MNRAS.371..983S}. These are constrained by $N_{\rm con}^{\rm arc} = 29\,469$ pixels within the arc mask (shown in magenta in Fig.~\ref{fig:cluster_multiple_images}). Due to the removal of some image families within the arc, the updated number of point-like constraints is $N_{\rm con}^{\rm img}=2\times84=168$ and the number of free point-like source parameters is $N_{\rm free}^{\rm src-pt}=2\times33+15=81$. Therefore, our complete lens model has $N_{\rm dof}^{\rm pos+ext} = (N^{\rm img}_{\rm con} + N_{\rm con}^{\rm arc}) - (N_{\rm free}^{\rm src-pt} + N_{\rm free}^{\rm src-ext}) = 29\,421$ degrees of freedom.

The tens of thousands of arc pixels dominate over the few hundred point-like images available. Consequently, the position-based likelihood term becomes subdominant compared to the extended-image likelihood term when evaluated close to the best-fit solution. Because we focus on the impact of modeling the extended surface of the SN H0pe host, we did not attempt to reweight the two likelihoods, so the extended-image term remains naturally dominant. We note, however, that Schuldt et al. (in prep.) explore different likelihood weighting choices, and find significant improvements in mass model uncertainties in all cases.

We ran Markov chain Monte Carlo chains to sample the lens mass parameters using the extended-image likelihood. We verified convergence using the power-spectrum method of \citet{Dunkley+2005}. Fig.~\ref{fig:corner_mass} shows the resulting mass parameter constraints as blue contours, which are substantially tighter than the image position constraints (red contours) from Sect.~\ref{ssec:SLmod:pointlike}. Averaging over all mass model parameters, including halo components and cluster members, we find that the posterior uncertainty is reduced by a factor of $\approx12$. This reduction is comparable to the gain reported by Schuldt et al. (in prep.) after including extended image modeling of the SN Refsdal host in their lens model of the cluster MACS\,J1149.5$+$2223.

\subsection{Other extended arcs and lens-model flexibility}

We performed extended-image modeling of the SN H0pe host (Arc 2), but did not reconstruct other arcs visible in Fig.~\ref{fig:cluster_multiple_images} for two reasons: (1) our work focuses primarily on the host of SN H0pe and (2) the computational costs would significantly increase. However, given the bimodality and configuration of \p, Arc 2 appears closer to cluster component 1. Therefore, we expect the parameters of component 1 to be better constrained, compared to those of component 2. Splitting the average gain over the mass parameters of components 1 and 2 yields factors of improvement of $\approx13$ and $\approx11$, respectively. Thus, Arc 2 provides slightly more constraints on cluster component 1. Modeling the extended surface brightness of arcs closer to cluster component 2 further reduces the uncertainty on all mass model parameters, even at the position of the lensed SN images. Nevertheless, it is not clear whether the uncertainty gain would benefit time-delay cosmography, given other sources of systematic errors and possible limitations of the mass model, as discussed below.

Our extended image modeling reveals residuals along the reconstructed Arc 2 (see Appendix~\ref{app:sec:source_all_bands} for details). We attribute part of these residuals, typically on the scale of a few pixels, to the specificity of the source reconstruction technique. However, residuals on the scale of multiple images, which are more prominent than small-scale residuals, may indicate missing flexibility in the lens model. Without adding additional components to the model, we could, for example, relax the scaling relation used to scale the mass parameters of cluster members (Eq.~\ref{eq:scaling_relation}) for galaxies near the modeled arc. For these galaxies, additional measurements of stellar velocity dispersions could complement the lensing constraints. Alternatively, we could introduce extra degrees of freedom in the cluster-scale components, such as multipoles or a variable density slope. We defer these improvements to future work.

\section{SN H0pe host galaxy \label{sec:sn_host}}

\subsection{Surface brightness reconstruction \label{sec:SN_host:surface_brightness}}

\begin{figure}
\centering
  \includegraphics[width = 0.8\columnwidth]{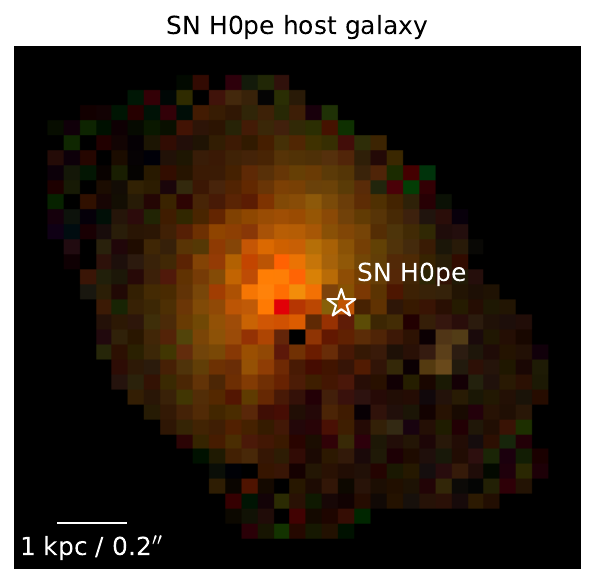}
  \caption{Color-composite of the reconstructed source-plane image of the SN H0pe host galaxy ($z=1.78$). The filters F090W, F150W, and F200W correspond to the blue, green, and red channels, respectively. The white star indicates the position of SN H0pe. The pixel scale is $0\farcs048$.}
  \label{fig:host_reconstruction}
\end{figure}

Our extended model of the SN H0pe host allowed \citet{2024ApJ...967...50P} to measure SN photometry and photometric time delays, which \citet{2025ApJ...979...13P} used to constrain $H_0$. Specifically, we used our Arc 2 model to subtract the host flux at each SN image position. Although we only used the F200W data to constrain the mass model parameters, we applied the best-fit lens model to reconstruct the extended source in other bands, which allowed \citet{2024ApJ...967...50P} to measure photometry in all filters bluer than F356W. Figure~3 of their paper shows the SN images after subtraction of our multi-band arc model; we refer interested readers to their paper for more details.

Appendix \ref{app:sec:source_all_bands} presents the complete set of source models. In the filter used for our lens model (F200W) and in redder filters, we observe residuals in high S/N areas in two of the three lensed images of the SN host galaxy (the middle and bottom images in the arc). Specifically, the middle image of the SN host shows positive residuals, indicating that our model under-magnifies the host, while the bottom image shows the opposite behavior. The top image, however, is significantly better reproduced by our lens model. These residuals may indicate that the two cluster-scale PIEMD components, particularly the one east of Arc 2, do not accurately represent the deflection field across the entire arc. Additionally, intra-cluster light (ICL) and light from low surface brightness cluster members or lower-redshift galaxy interlopers below the detection limit may also explain the excess light in the vicinity of Arc 2b \citep[see e.g.,][]{2022ApJ...932...85P}. We defer Including the surface brightness model of foreground objects to a future work.

Figure ~\ref{fig:host_reconstruction} shows the reconstructed color image of the SN host.  Despite the low S/N in the bluer filters (especially F090W), we observe a clear color gradient: the core of the SN H0pe host galaxy appears redder than its outskirts. Figure\ref{fig:host_reconstruction} shows that SN H0pe lies about 1~kpc from its host centroid, in a region slightly darker than other regions at similar galacto-centric radii. These darker source regions may be attenuated by dust, which we investigate next.

\subsection{Dust extinction in the image plane \label{ssec:dust:image_plane}}

Dust lanes appear in the SN host images, typically in the bluer filters of the set \citep[e.g., the redder regions in Fig.~1 of][]{2025ApJ...979...13P}. Dust can affect photometric measurements, especially the observed brightness of the SN images, potentially biasing the time delays used in cosmographic analyses. Recent JWST studies have measured the dust properties of the SN H0pe host. \citet{2024ApJ...961..171F} measured quantities such as the color excess $E(B-V)$ using a NIRCam+NIRSpec spectrophotometric analysis, \citet{2024ApJ...970..102C} fit NIRSpec spectra at the position of each image to infer both the extinction ratio $R(V)$ and $E(V-B)$, and \citet{2024ApJ...967...50P} constrained the total V-band dust extinction, $A_V \equiv R(V) \times E(B-V)$ jointly with $R(V)$ and the time delays. Complementing these approaches, we constructed a dust extinction map of the SN H0pe host by mapping the spatial distribution of $A_V$ across the entire Arc 2.

We followed the procedure of \citet[in particular their Sect.~5.4]{2009ApJ...691..277S} to build a dust extinction map from the NIRCam images. We first express the observed magnitude $m_{F,\rm{observed}}$ in a given filter $F$ in
terms of $A_V$ and the intrinsic magnitude of the reddest wavelength band $m_{1,\rm{intrinsic}}$ as
\begin{align}
\label{eq:mobsInAvm1}
m_F \equiv m_{F,\rm{observed}} =  m_{1,\rm{intrinsic}} + Q_F + A_V k_F + n_F \ ,
\end{align}
where $k_F \equiv {A_{F}}/{A_V}$ are constants given by the extinction law and $n_F$ is the noise in band $F$. We define $Q_F$ as the intrinsic color relative to the reddest filter:
\begin{align}
\label{eq:q_f}
    Q_F = m_{F,\rm{intrinsic}}-m_{1,\rm{intrinsic}} \ ,
\end{align}
where $F=1,\ldots, N_{\rm{b}}$ is the sequence of $N_{\rm{b}}$ filters ordered from the reddest to the bluest (by construction $Q_{1}=0$). We converted the original flux data units (MJy/st) to AB magnitudes using zero-point values of 28.0874 (blue arm) and 28.0863 (red arm) \footnote{This differs slightly from the zero-point value in \citet[that is 28.0865]{Windhorst2023}, which we attribute to the slightly different reduction of \citet{2024ApJ...961..171F}, who used calibration files available at that time.}.

\citet{2009ApJ...691..277S} showed that $A_{V}$ and $m_{1,\rm{intrinsic}}$ can be solved jointly. Here, we focus on $A_{V}$, which has the following analytical solution:
\begin{eqnarray}
\label{eq:a_v}
A_V =& \bigg[ &\frac{1}{N_{\rm{b}}}\left(\sum_F k_F \right) \left(\sum_F m_F \right) - \nonumber \\
& & - \frac{1}{N_{\rm{b}}}\left(\sum_F k_F \right) \left(\sum_F Q_F \right) - \nonumber \\
& & - \sum_F k_F m_F + \sum_F k_F Q_F \ \ \ \bigg]\ \ \ \bigg/ \nonumber \\
& \bigg[& \frac{1}{N_{\rm{b}}} \left(\sum_F k_F\right)^2 - \sum_F k_F^2 \ \ \ \bigg] \ .
\end{eqnarray}

To compute $A_V$ we estimated the intrinsic colors $Q_F$ from Eq.~\ref{eq:q_f}. We used F200W, F270W, F356W, and F444W, with F444W being the reddest and thus our reference filter (i.e., with $F=1$). We excluded NIRCam filters bluer than F200W because the arc S/N is too low, and noise patterns introduce artifacts in the dust extinction map (see Fig.~\ref{app:fig:all_bands}). Because bluer filters have intrinsically higher resolution than redder ones, we re-convolved all filters to match the resolution of our reference filter. We constructed resolution-matching PSF kernels using the \texttt{create\_matching\_kernel()} routine (using a \texttt{CosineBellWindow} regularizing function to avoid artifacts) from the \textsc{photutils} Python package \citep{larry_bradley_2024_13989456}. During this process, we found that PSF models created with \textsc{starred} \citep{Michalewicz2023,2024AJ....168...55M}---instead of the stacked-stars models used for lens modeling---improved both the resolution-matching step and the resulting dust extinction map results\footnote{Using these alternative PSF models does not significantly affect the lens modeling results in Sect.~\ref{ssec:SLmod:esr}.}.

\begin{figure*}
  \includegraphics[width=\linewidth]{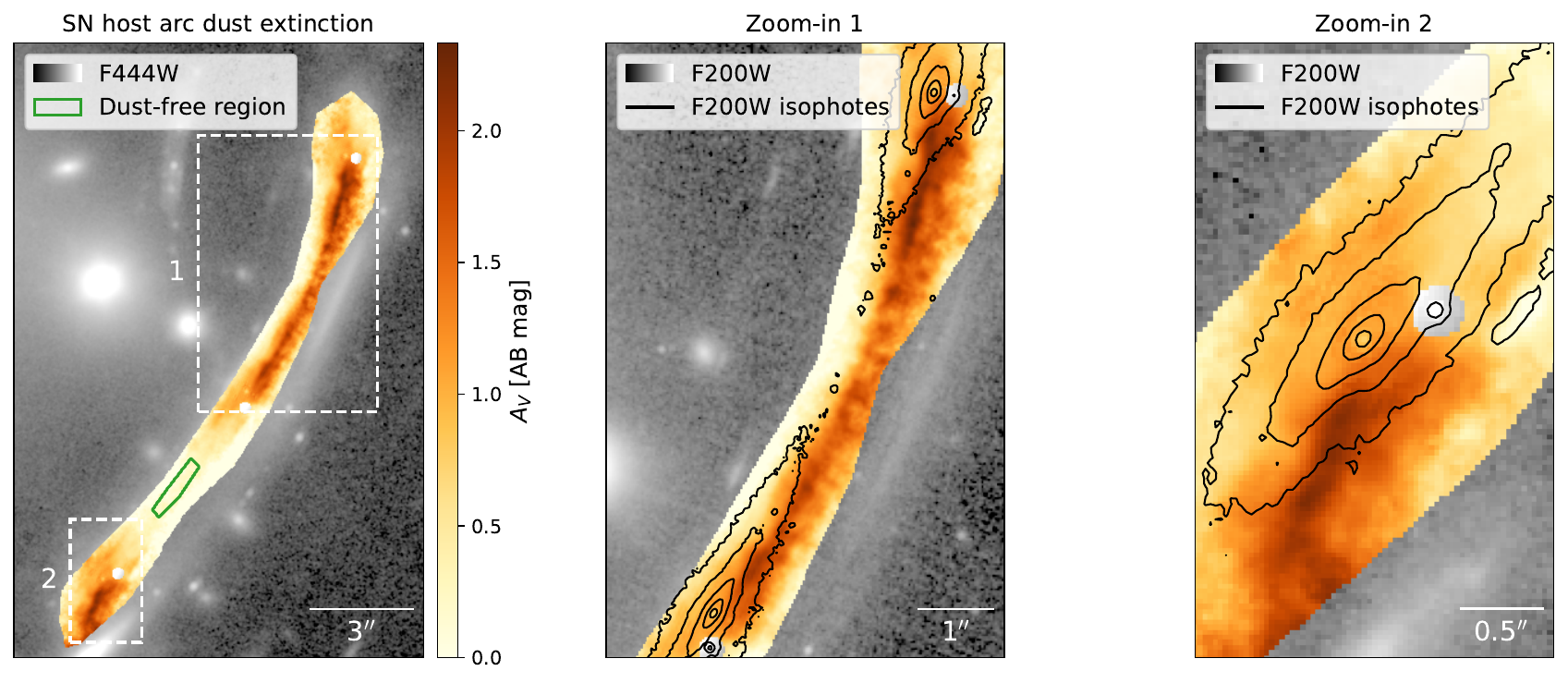}
  \caption{Image-plane dust extinction map of the SN H0pe host galaxy. Left panel: Dust extinction map within the arc mask, at the resolution of the F444W data, which is also shown in grayscale (see Sect.~\ref{ssec:dust:image_plane} for details). The dashed-line rectangles indicate the zoom-in regions of the middle and right panels. Middle and right panels: F200W data in grayscale and corresponding isophotes (logarithmically spaced), showing consistency between dust extinction and dimmer regions of the arc. In all panels, north is at the top and east is to the left.}
  \label{fig:dust_map_image_plane}
\end{figure*}

We visually identified a region in the arc that is likely not affected by dust (shown on the left of Fig.~\ref{fig:dust_map_image_plane}). Within this region, we  carefully selected a smaller region with the bluest color relative to F444W. We restricted the region in two steps. First, we applied a $5\sigma$ noise threshold simultaneously in both filters for a given pair (Eq.~\ref{eq:q_f}). We then adopted the first percentile of the resulting distribution as our final color estimate, i.e., the pixel that is bluer than 99\% of the pixels within the restricted region. This procedure excludes pixels that are not representative of the bulk distribution of colors across our region. Such outlier pixels may appear too blue due to noise fluctuations or localized bright features, such as star-forming regions, which may belong to the foreground or the host galaxy.

After estimated the intrinsic colors $Q_F$, we evaluated Eq.~\ref{eq:a_v} for every pixel of the arc, adopting the reddening law of \citet{1999PASP..111...63F} \footnote{Using the extinction law of \citet{1989ApJ...345..245C} produces negligible differences. We used the Python package \texttt{\href{https://github.com/sncosmo/extinction}{extinction}} for its implementation of various extinction laws.} and extinction ratio $R_V = 3.1$ (i.e., galactic extinction), again following \citet{2009ApJ...691..277S}. Fewer than $10\%$ of the pixels yield slightly negative $A_V$ values, which is unphysical. These pixels generally lie near the edge of the arc mask, where flux from the ICL or other galaxies, low S/N, or slight over-subtraction of the background can produce negative fluxes. Figure~\ref{fig:dust_map_image_plane} shows the map of dust extinction $A_V$ for each pixel within the arc mask, excluding pixels around the SN images. The middle and right panels of Fig.~\ref{fig:dust_map_image_plane} confirm that areas with higher dust extinction correspond to lower-intensity regions of the arc in bluer filters (for example, in F200W; the trend is also clear in F150W despite its lower S/N).

The $A_V$ map excludes circular regions centered on each of the SN images\footnote{These regions match those excluded from the extended source model likelihood in Sect.~\ref{ssec:SLmod:esr}.}. To estimate the dust extinction at the position of SN H0pe, we interpolated the $A_V$ values within the masked regions. We used nearest-neighbor interpolation, which relies on minimal assumptions regarding the underlying $A_V$ spatial variations and is accurate enough for the few pixels involved. In principle, the dust extinction at the position of each SN image should be identical, but factors such as data noise, PSF modeling (original and resolution-matching kernels), and light contamination from foreground objects introduce differences. We therefore adopt the mean over the three images as our fiducial value, for which we find $A_{V}^{\rm H0pe} = 0.94 \pm 0.25$. We computed the uncertainty as the quadratic sum of three error terms. For the statistical error term, we emulated 100 realizations of the data and took the resulting standard deviation of $A_V$, averaged over the three SN image positions. We then estimated a systematic error from the PSF model choice by taking the mean difference between the $A_{V}^{\rm H0pe}$ values obtained with the PSF models used in lens modeling and those obtained using \textsc{starred}. Our final error term is the standard deviation over the $A_V$ values interpolated at each position of the SN images (we used the Bessel correction for the standard deviation due to the small sample size). These three error terms due to data noise, PSF model, and image position account for $1\%$, $6\%$, and $93\%$ of the total variance, respectively.

\subsection{Dust extinction in the source plane \label{sssec:dust:source_plane}}

\begin{figure}
\centering
  \includegraphics[width = 0.8\columnwidth]{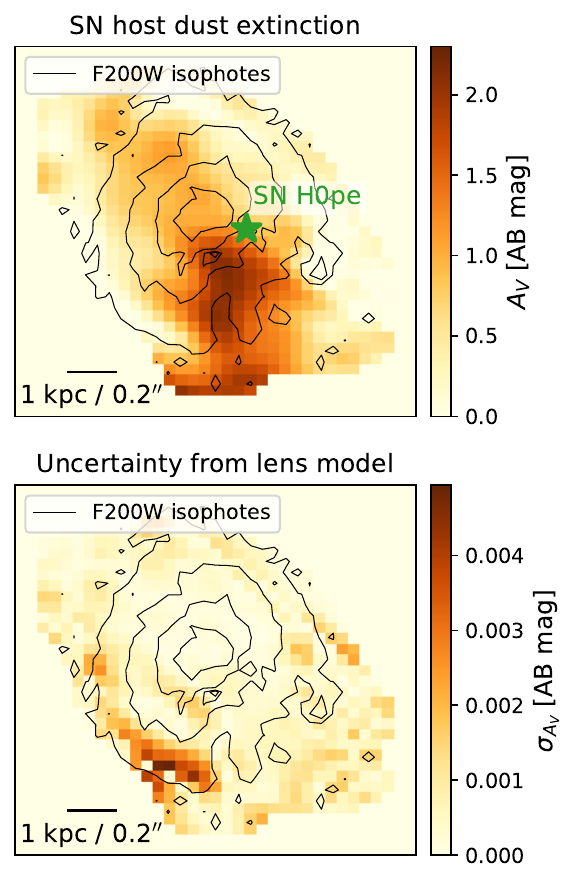}
  \caption{Top panel: Dust extinction map of the SN H0pe host galaxy, reconstructed in the source plane based on our extended-image lens model. The green star marks the SN position relative to its host; thin black contours show its (logarithmically spaced) isophotes. We note the consistency between the perturbed isophotes and the dust extinction map. The bottommost high-extinction feature (about three bottom rows of the source-plane grid) arises mainly from low data S/N and image-plane flux contamination at the edge of the arc mask. North is at the top and east is to the left. Bottom panel: Uncertainty in $A_V$ due to the posterior uncertainty of our strong lensing model. In both panels, the pixel scale is $0\farcs048$.}
  \label{fig:dust_map_source_plane}
\end{figure}

We used the dust extinction map from Fig.~\ref{fig:dust_map_image_plane} and reconstructed it on the source plane using our lens model from Sect.~\ref{ssec:SLmod:esr}. Because the \glee source reconstruction algorithm requires an estimate of the uncertainty per pixel, we approximated it by $\sqrt{A_V}$ (i.e., similar to a shot noise estimate), which results in a reasonable source-plane reconstruction. The top panel of Fig.~\ref{fig:dust_map_source_plane} presents the source-plane dust extinction map of the SN H0pe host galaxy. SN H0pe lies on the outskirts of a large region with significant dust extinction, with $A_V\gtrsim1$. We did not derive an interpolated $A_V^{\rm H0pe}$ value from our source-plane dust extinction map. Such an estimate would be too uncertain due to the imperfect PSF model (whose effect is more pronounced than in the image plane), pixelation effects in the source reconstruction, and the specific per-pixel $A_V$ uncertainty. Instead, our image-plane estimate detailed in Sect.~\ref{ssec:dust:image_plane} is more robust and should be preferred.

We find that the dust distribution in the SN host is strongly asymmetric. Whereas early-type galaxies and mid-type spirals generally display symmetric and radially decreasing $A_V$ profiles \citep[see e.g.,][]{2019ApJ...884...21K}, the host of SN H0pe clearly shows stronger reddening south of the galaxy center. This dust pattern could result from outflow mechanisms that triggered a burst of star formation. Alternatively, inflows of dusty material could explain the asymmetric feature visible on the dust extinction map. Dedicated analyses of the available NIRSpec spectra from each lensed image could provide insights, although additional observations at submillimeter wavelengths would certainly help to clarify the origin of the asymmetric dust distribution.

We also observe a relatively high-extinction region at the bottom of the source-plane grid in Fig.~\ref{fig:dust_map_source_plane}. This region maps to the edge of the arc mask in the bottom lensed image (right panel of Fig.~\ref{fig:dust_map_image_plane}) and primarily arises from low S/N in the bluer filters, as well as possible contamination from the neighboring arc (see Sect.~\ref{ssec:colors_arc7}). To assess if this high-extinction region has high uncertainty, we computed the $A_V$ standard deviation in each pixel of the source plane from 100 random lens model posterior samples. The bottom panel of Fig.~\ref{fig:dust_map_source_plane} shows the resulting uncertainty map. The map indicates that the bottom-most high-extinction region is not particularly uncertain based on our lens model, suggesting it likely represents a real feature of the image-plane dust distribution. We did not construct a complete uncertainty map that includes other error sources, since these are more challenging to properly propagate to the source plane.

\subsection{Comparison with previous work and influence on $H_0$ \label{sssec:dust:av_comparison}}

\citet{2024ApJ...967...50P} inferred the dust extinction at the SN H0pe position as a by-product of their photometric time delay measurement. They performed a spectral energy distribution (SED) fitting with a free parameter, including $R_V$ and $A_V$ at the SN H0pe position, and inferred $A_{V}^{\rm H0pe,P24} = 1.21\pm0.11$. This value agrees reasonably well with our measurement, at the $\sim1\sigma$ level. This agreement is noteworthy because the two measurements are independent and use some of the same data but in different ways. \citet{2024ApJ...967...50P} performed PSF photometry on host-subtracted images based on our extended host model from Sect.~\ref{ssec:SLmod:esr}, using only pixels within small regions around each SN image. In contrast, we excluded these regions to build our dust extinction map and estimated $A_{V}^{\rm H0pe}$ through interpolation. Despite these differences, the approaches are complementary and both yield similar dust extinction values.

\citet{2024ApJ...970..102C} also inferred dust parameters simultaneously with the spectroscopic time delays of SN H0pe. They fit the NIRSpec spectra of each image using a set of shared parameters, notably $R_V$ and $E(B-V)$. They obtained $R(V)=2.7_{-0.1}^{+0.2}$ and $E(B-V)=0.27\pm0.01$, resulting in a total extinction of $A_{V}^{\rm H0pe,C24} = 0.73^{+0.08}_{-0.05}$. Although their result is unsurprisingly more precise than ours (since they use a high-resolution spectral fitting), both results are in good agreement at the $\sim0.8\sigma$ level. Again, it is reassuring that very different approaches, either based solely on NIRCam photometry or NIRSpec spectroscopy, produce consistent measurements. More recently, \citet{2026MNRAS.tmp..317G} remeasured the time delays between SN H0pe images by performing SED modeling of the photometric light curves from \citet{2024ApJ...967...50P}, including microlensing effects in their model. In addition to the SN time of maximum, they measured $A_{V}^{\rm H0pe,G25} = 0.95\pm0.14$. This value again agrees well with our measurements.

Although not exactly at the SN H0pe position, \citet{2024ApJ...961..171F} measured the color excess $E(B-V)$ for Arcs 2a and 2c using NIRCam+NIRSpec spectrophotometric SED fitting. Their extracted NIRSpec spectra do not contain the SN H0pe images, but rather regions near the host galaxy center (top panel of their Fig.~4). They measured $E(B-V)=0.22\pm0.04$ for Arc 2a and $E(B-V)=0.20\pm0.04$ for Arcs 2c. Assuming $R(V)=3.1$ (as in Sect.~\ref{ssec:dust:image_plane}), these measurements correspond to $A_V^{\rm 2a,F24} = 0.68\pm0.1$ and $A_V^{\rm 2c,F24} = 0.62\pm0.1$.  The top panel of fig.~\ref{fig:dust_map_source_plane} shows $A_V\lesssim1$ within the central isophote of our source-plane dust extinction map, which is broadly consistent with their measurements.

\citet{2025ApJ...979...13P} used the photometric time delays of SN H0pe measured in \citet{2024ApJ...967...50P} with our extended arc model for cosmography. Therefore, we can qualitatively discuss how dust extinction may influence the inferred value of $H_0$. Fig.~5 of \citet{2024ApJ...967...50P} shows mild correlations between the two time delays and $A_V$, which are slightly more pronounced for $\Delta t_{\rm ba}$ between images 2a and 2b of SN H0pe (top and middle images; see Fig.~\ref{fig:cluster_multiple_images}). Specifically, $A_V$ and $\Delta t_{\rm ba}$ are anti-correlated. Our lower value of $A_{V}^{\rm H0pe}$ relative to the best-fit of \citet{2024ApJ...967...50P} would slightly increase $\Delta t_{\rm ba}$ by less than ten days \citep[which is within the 1$\sigma$ uncertainty of $\Delta t_{\rm ba}$  measured by][]{2024ApJ...967...50P}.  Because $\Delta t_{\rm ba}$ is negative, this increase makes $\Delta t_{\rm ba}$ less negative, with a lower absolute value. Qualitatively, a lower absolute value of the time delay $|\Delta t|$ leads to a higher value $H_0$ \citep[this can also be deduced from Fig.~3 of][]{2025ApJ...979...13P}. Accurately quantifying the true impact on $H_0$ is nontrivial, as the final measurement of \citet{2025ApJ...979...13P} is the result of a thorough Bayesian analysis combining multiple underlying models. Furthermore, the covariance between $A_V$ and $\Delta t_{\rm ba}$ would need to be carefully reassessed, as it may differ from \citet{2024ApJ...967...50P} given the many differences between our methodologies. Nevertheless, because our $A_V$ values agree reasonably well with those of \citet{2024ApJ...967...50P} and our work, and because the effect on the measured time delay is small a priori, the impact on $H_0$ is expected to be subdominant with respect to the current error budget.

\subsection{Intrinsic colors of Arc 2 and dust in Arc 7 \label{ssec:colors_arc7}}

As detailed in Sect.~\ref{ssec:dust:image_plane}, our dust extinction procedure uses an estimation of the SN host intrinsic colors. We carefully selected a specific region in Arc 2 that is likely unaffected by dust and assumed it  represents the intrinsic color of the SN host galaxy. Although this assumption holds for a symmetric, smooth galaxy such as an elliptical, it may not be accurate for one with more complex morphology. Smooth variations in the intrinsic color may also arise from local changes in stellar metallicities and stellar populations. The joint NIRCam and NIRSpec analysis of two of the three host images, presented in detail by \citet{2024ApJ...961..171F}, indicates that it is a moderately dusty and massive star-forming galaxy, with a possibly complex star formation history. Our dust extinction map corroborates such conclusions. The low S/N in bluer filters as well as the light contamination from nearby arcs---including galaxies with very similar colors, belonging to the same $z=1.78$ group as the SN host \citep{2024ApJ...961..171F}---make it challenging to obtain better estimates of the intrinsic colors. The estimate presented here carefully excluded regions that were either too low in S/N or whose colors could be affected by local features (e.g., SN images or compact star-forming regions). An alternative approach to ours would be to perform SED fitting for each pixel within the arc mask. Such an approach would remove the need to assume intrinsic colors that do not vary over the scale of the galaxy. However, in addition to being computationally more expensive, per-pixel SED fitting might be limited by the relatively low number of JWST bands available, due to known degeneracies among dust extinction, stellar populations, and metallically.

We note that the SN H0pe host galaxy (Arc 2) lies very close to another equally extended arc (referred to as Arc 7 and indicated in Fig.~\ref{fig:cluster_multiple_images}). From lensing geometry, one can argue that Arc 7 corresponds to a larger Einstein radius than Arc 2, hence further away from us. However, other arguments using symmetry points within each arc could support the opposite scenario. If Arc 7 is indeed closer to us and contains a significant amount of dust, then it could be responsible for at least some of the dimming and reddening visible along Arc 2. To the best of our knowledge, there is currently no spectroscopic redshift for Arc 7. \citet{2019ApJ...871...51F} estimate $z=1.7$ from their lens models, and \citet{2022ApJ...932...85P} report a photometric redshift of $z=1.86_{-0.28}^{+0.29}$. In all our models, the redshift of Arc 7 is left as a free parameter, but given the low number of secure point-like constraints associated with it (3), this redshift is not robustly constrained. Therefore, one cannot currently rule out the possibility that Arc 7 contributes to the observed dust extinction. We emphasize that our image-plane dust extinction map (Fig.~\ref{fig:dust_map_image_plane}) is unaffected by the true distance of Arc 7 relative to Arc 2, since we do not make any assumptions regarding the origin of the dust. On the contrary, our source-plane dust extinction map (top panel of Fig.~\ref{fig:dust_map_source_plane}) could change with knowledge of the redshift of Arc 7. For instance, the southernmost area with significant $A_V$ values in our source-plane extinction may be the result of dust from another lensed galaxy, which could be Arc 7 if it is located in the foreground.

\section{Conclusions}
\label{sec:conclusions}

The full exploitation of strongly lensed Type Ia SNe such as SN H0pe relies on a good lens model. In this work, we show a significant gain in the precision of the mass model parameters after incorporating all the pixels from the arc hosting SN H0pe. We performed two lens models of the host cluster \p: one using the traditional position-based approach with point-like multiple images as constraints, and an improved model that jointly reconstructs the full surface brightness of the arc. We find that including the extended SN host as a constraint decreases the uncertainty on mass model parameters by a factor of $\gtrsim10$. This gain in precision would translate to improved precision in the Hubble constant, which must be quantified by following the approach of \citet{2025ApJ...979...13P}.

Our improved lens model, the first to reconstruct the surface brightness of the SN H0pe host galaxy, enables the study of the dust distribution in the galaxy. In particular, we computed the dust extinction map over the entire gravitational arc, leveraging the color information in the JWST/NIRCam dataset. We find good agreement between the visibly attenuated regions in the arc and those with high values of the extinction coefficient. Our lens model maps the dust extinction back to the plane of the host galaxy, showing that SN H0pe lies on the edge of a dust region. The relatively complex yet moderate dust extinction across the host is qualitatively consistent with the conclusions of \citet{2024ApJ...961..171F}, based on simultaneous fitting of the JWST/NIRCam imaging and JWST/NIRSpec spectroscopy of the arc. Our dust extinction map yields a V-band extinction of $A_{V}^{\rm H0pe} = 0.94 \pm 0.25$ at the position of SN H0pe. Our estimate is in good statistical agreement with the fully independent value of \citet{2024ApJ...967...50P}, based on multi-epoch photometry, and with that of \citet{2024ApJ...970..102C}, based on multi-epoch spectroscopy.

Our direct modeling of the SN host surface brightness reveals model residuals along the arc that point to limitations in the parametric model of \p. In a follow-up work (Galan et al., in prep.), we will improve our lens models by including foreground light from the brightest cluster members and the ICL, and by increasing the flexibility of both the lens and source models to better fit the observations. We will use the approach introduced in \citet{2024A&A...689A.304G}, employing the lens modeling code \textsc{Herculens} \citep{2022A&A...668A.155G}. Comparing our present and future lens models with those used in \citet{2025ApJ...979...13P} will enable further investigation of the systematic biases reported in \citet{Agrawal2025}, ultimately improving the measurement of $H_0$ with time-delay cosmography on cluster scales. Moreover, newly obtained JWST/NIRCam template images will, together with our results on the spatially resolved dust extinction, enable improved photometry and time delays from SN H0pe.

\begin{acknowledgements}
SS has received funding from the European Union’s Horizon 2022 research and innovation programme under the Marie Skłodowska-Curie grant agreement No 101105167 — FASTIDIoUS.  
SHS and SE thank the Max Planck Society for support through the Max Planck Fellowship for SHS. This project has received funding from the European Research Council (ERC) under the European Union's Horizon 2020 research and innovation programme (LENSNOVA: grant agreement No 771776).
This work is supported in part by the Deutsche Forschungsgemeinschaft (DFG, German Research Foundation) under Germany's Excellence Strategy -- EXC-2094 -- 390783311. 
RAW acknowledges support from NASA JWST Interdisciplinary Scientist grants
NAG5-12460, NNX14AN10G and 80NSSC18K0200 from GSFC.
CG acknowledges financial support through grant MIUR2020 SKSTHZ.
AA acknowledges financial support through the Beatriz Galindo programme and the project PID2022-138896NB-C51 (MCIU/AEI/MINECO/FEDER, UE), Ministerio de Ciencia, Investigación y Universidades.
J.M.D. acknowledges the support of projects PID2022-138896NB-C51 (MCIU/AEI/MINECO/FEDER, UE) Ministerio de Ciencia, Investigaci\'on y Universidades and SA101P24 (Junta de Castilla y Le\'on).
RAW acknowledges support from NASA JWST Interdisciplinary Scientist grants NAG5-12460, NNX14AN10G and 80NSSC18K0200 from GSFC.
This work is based in part on observations made with the NASA/ESA/CSA James Webb Space Telescope. The data were obtained from the Mikulski Archive for Space Telescopes at the Space Telescope Science Institute, which is operated by the Association of Universities for Research in Astronomy, Inc., under NASA contract NAS 5-03127 for JWST. These observations are associated with program 1176. The data were obtained from the Barbara A. Mikulski Archive for Space Telescopes (MAST) at the STScI (accessible via \url{https://dx.doi.org/10.17909/zk1p-2q51}), which is operated by the Association of Universities for Research in Astronomy (AURA) Inc., under NASA contract NAS 5-26555.
This research made use of \textsc{SciPy} \citep{Virtanen2020scipy}, \textsc{NumPy} \citep{Oliphant2006numpy,VanDerWalt2011numpy}, \textsc{Matplotlib} \citep{Hunter2007matplotlib}, \textsc{Astropy} \citep{astropy2013,astropy2018} and \textsc{GetDist} \citep{Lewis2019getdist}.
\end{acknowledgements}

\bibliographystyle{aa}
\bibliography{references}

\begin{appendix}

\onecolumn

\section{Full joint posterior distribution}

\begin{figure*}[ht!]
\centering
  \includegraphics[width = 1.\textwidth]{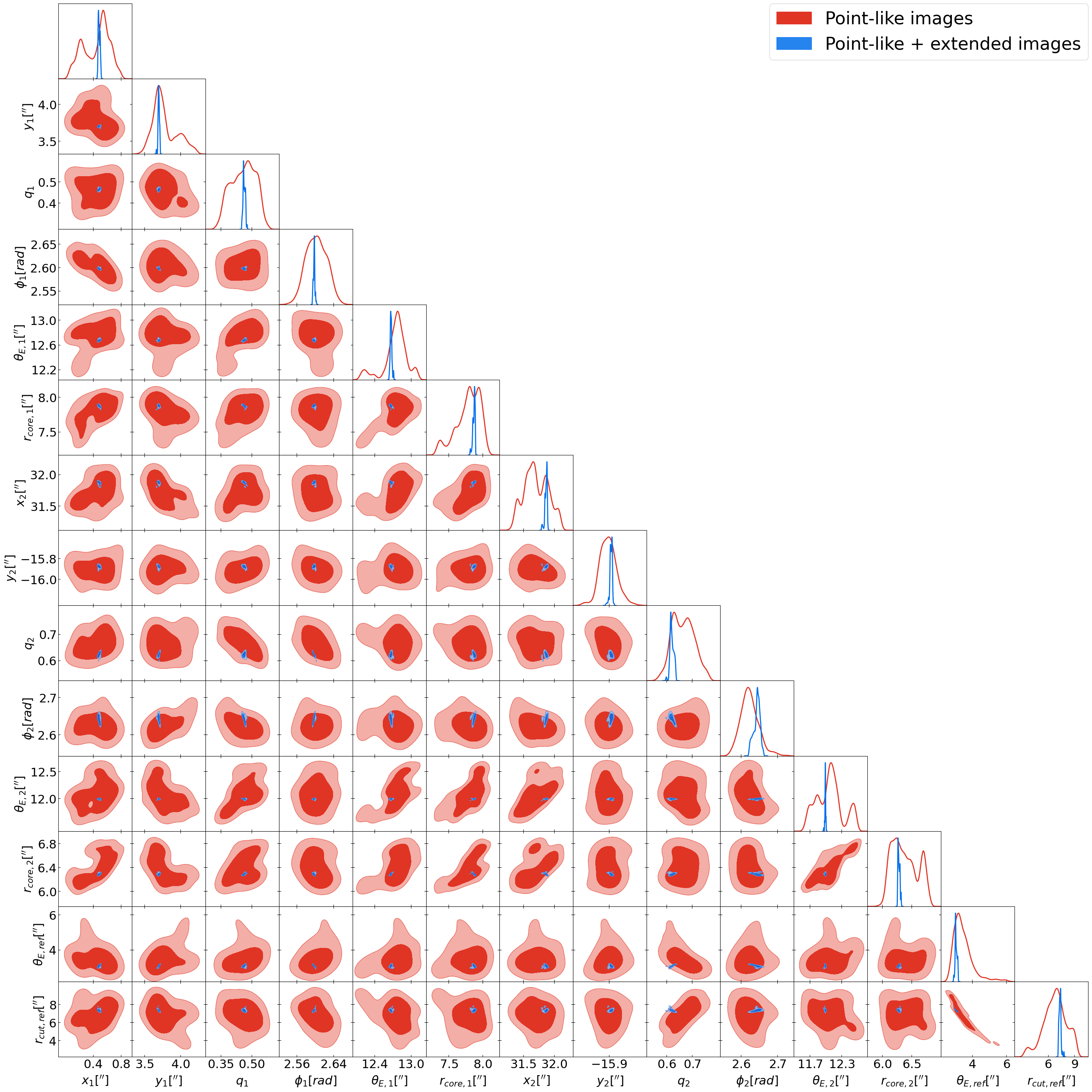}
  \caption{Corner plot for all mass parameters for the two cluster scale components and scaling relations of the cluster members. Contours are the 68\% and 95\% confidence levels for the model using only the positions of the point-like multiple images in red and the full surface brightness of SN H0pe host galaxy in blue.}
  \label{app:fig:corner}
\end{figure*}

We show in Fig.~\ref{app:fig:corner} all mass model parameters sampled for our lens models. All parameters are significantly better constrained by including the extended surface brightness of SN H0pe host galaxy.

\section{Reconstructed SN host in multiple JWST bands \label{app:sec:source_all_bands}}

\begin{figure*}[ht!]
\centering
  \includegraphics[width=0.62\linewidth]{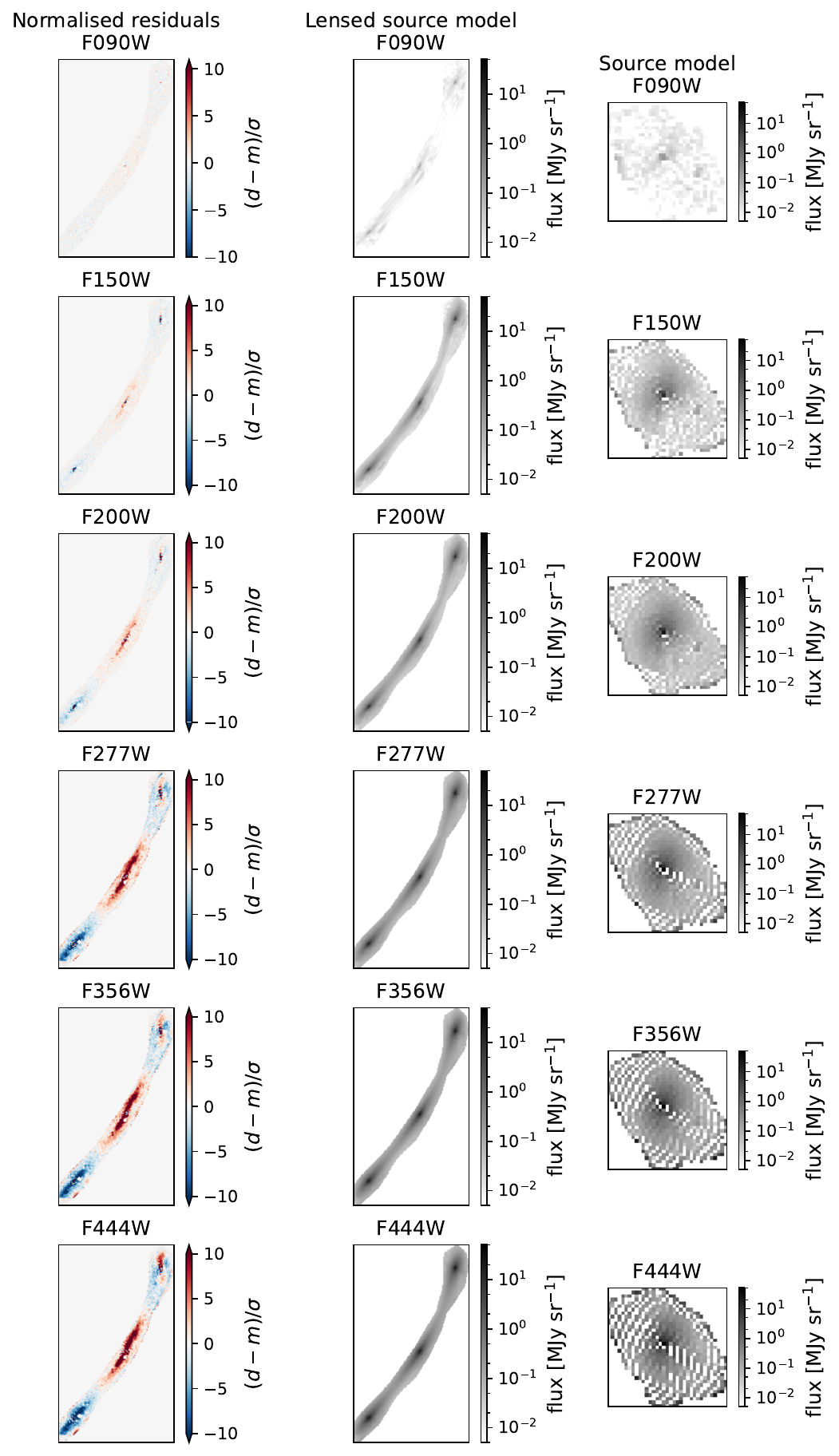}
  \caption{Extended SN H0pe host reconstruction in six JWST/NIRCam bands. Our lens model parameters are constrained using F200W only (third row). Subsequently, other bands were used to reconstruct the source surface brightness, fixing the best-fit mass parameters. Normalized residuals (left column) are defined as data $-$ model divided by the uncertainty in each pixel.}
  \label{app:fig:all_bands}
\end{figure*}

In Sect.~\ref{ssec:SLmod:esr} we obtained an improved lens model using the F200W data and including the extended SN H0pe arc as extra constraints. In Fig.~\ref{app:fig:all_bands} we show the reconstructed extended host surface brightness in six NIRCam filters, fixing the lens model to the best-fit parameters found with F200W.

We note that while the arc is fitted reasonably well in all bands, especially in bluer ones, large scale patterns in the residuals still remain. Part of these residuals are caused by foreground light contamination from cluster members, in particular from the brightest galaxy at the top left of Fig.~\ref{fig:cluster_multiple_images}, which is not included in our model. A second origin of residuals may be due to the mass model being too simplistic, which does not accurately reproduce the magnification of the two southernmost images of the arc (one appears over-magnified, while the other seems under-magnified). Both of these effects---foreground light and mass distribution---will be the focus of a future paper using the modeling approach of \citet{2024A&A...689A.304G}.

\end{appendix}

\end{document}